\documentclass[
aps,
prl,
showpacs
amsmath,
amssymb,
floatfix,
longbibliography,
letterpaper,
lengthcheck,
superscriptaddress,
nofootinbib
]{revtex4-1}

\usepackage{graphicx}
\usepackage{amsmath}
\usepackage{braket}

\newcommand{\myTitle}{Cavity Born-Oppenheimer Approximation for Correlated Electron-Nuclear-Photon Systems}

\begin{document}

\title{\myTitle}

\author{Johannes Flick}
  \email[Electronic address:\;]{johannes.flick@mpsd.mpg.de}
  \affiliation{Max Planck Institute for the Structure and Dynamics of Matter and Center for Free-Electron Laser Science, Department of Physics, Luruper Chaussee 149, 22761 Hamburg, Germany}
\author{Heiko Appel}
  \email[Electronic address:\;]{heiko.appel@mpsd.mpg.de}
  \affiliation{Max Planck Institute for the Structure and Dynamics of Matter and Center for Free-Electron Laser Science, Department of Physics, Luruper Chaussee 149, 22761 Hamburg, Germany}
\author{Michael Ruggenthaler}
  \email[Electronic address:\;]{michael.ruggenthaler@mpsd.mpg.de}
  \affiliation{Max Planck Institute for the Structure and Dynamics of Matter and Center for Free-Electron Laser Science, Department of Physics, Luruper Chaussee 149, 22761 Hamburg, Germany}
\author{Angel Rubio}
  \email[Electronic address:\;]{angel.rubio@mpsd.mpg.de}
  \affiliation{Max Planck Institute for the Structure and Dynamics of Matter and Center for Free-Electron Laser Science, Department of Physics, Luruper Chaussee 149, 22761 Hamburg, Germany}
  \affiliation{Nano-Bio Spectroscopy Group and ETSF,  Dpto. Fisica de Materiales, Universidad del Pa\'is Vasco, 20018 San Sebasti\'an, Spain}

\date{\today}

\begin{abstract}
In this work, we illustrate the recently introduced concept of the cavity Born-Oppenheimer approximation for correlated electron-nuclear-photon problems in detail. We demonstrate how an expansion in terms of conditional electronic and photon-nuclear wave functions accurately describes eigenstates of strongly correlated light-matter systems. For a GaAs quantum ring model in resonance with a photon mode we highlight how the ground-state electronic potential-energy surface changes the usual harmonic potential of the free photon mode to a dressed mode with a double-well structure. This change is accompanied by a splitting of the electronic ground-state density. For a model where the photon mode is in resonance with a vibrational transition, we observe in the excited-state electronic potential-energy surface a splitting from a single minimum to a double minimum. Furthermore, for a time-dependent setup, we show how the dynamics in correlated light-matter systems can be understood in terms of population transfer between potential energy surfaces. This work at the interface of quantum chemistry and quantum optics paves the way for the full ab-initio description of matter-photon systems.
\end{abstract}
 
\date{\today}

\maketitle


\section{Introduction}
\label{sec:Intro}
Recent experimental progress has made it possible to study light-matter interactions in the regime of strong and ultra-strong light-matter coupling. Experiments from exciton condensates~\cite{vasilevskiy2016,low2016}, near-field spectroscopy, plasmon mediated single molecule strong coupling~\cite{chikkaraddy2016}, superconducting qubit circuits~\cite{diaz2016}, quantum information~\cite{blais2004}, direct measurements of vacuum fluctuations in solids~\cite{riek2015}, and chemistry in optical cavities~\cite{shalabney2015,orgiu2015,george2016}
open now the path to shape the emerging field-fluctuations with the goal towards a new control of material properties. In this new field that has been driven in particular by experiment, traditional theoretical methods from either quantum chemistry or quantum optics loose their applicability. On the one hand, traditional quantum chemistry concepts such as the Born-Oppenheimer (BO) approximation~\cite{born1927,gross1991}, {or electronic structure methods} such as Hartree-Fock theory~\cite{szabo1989}, coupled-cluster theory~\cite{bartlett2007}, or density-functional theory (DFT)~\cite{hohenberg1964} have been originally designed to treat approximately correlated electron-nuclear problems but are not capable to correctly {account} for the quantum nature of light. On the other hand, concepts from quantum optics typically describe the quantum nature of the light field in great detail, but fail in describing more complex dynamics of matter due to {the often employed} simplification to {a} few levels~\cite{fleischhauer2000,loudon2000}. To fill this gap, in this work, we generalize a well-established concept from quantum chemistry, namely the Born-Oppenheimer approximation, to the realm of correlated light-matter interactions for systems in optical high-Q cavities.\\
First theoretical studies in similar direction, e.g. the modification of the molecular structure under strong light-matter coupling~\cite{galego2015}, the nonadiabatic dynamics of molecules in optical cavities~\cite{kowalewski2016,kowalewski2016b}, or the cavity-controlled chemistry~\cite{herrera2016} have already been conducted.\\
Since the complexity of an exact ab-initio description of such correlated many-body systems that contain electronic, nuclear, and photonic (fermionic and bosonic) degrees of freedom scales exponentially with system size, approximate descriptions have to be employed for any realistic system. Recently, the concept of DFT has been generalized to electron-photon problems and was termed quantum-electrodynamical density-functional theory (QEDFT)~\cite{ruggenthaler2011b,tokatly2013,ruggenthaler2014,ruggenthaler2015}. This theory maps the complicated many-body problem into a set of nonlinear equations for the electronic and photonic degrees of the densities/currents that {facilitates} the treatment of such complex systems, similarly as standard DFT has done over the years to deal with correlated electronic systems. Still for this theory to be applicable{,} accurate functionals for combined light-matter systems have to be developed to calculate approximate effective potentials and observables. In this work, we use an alternative approach, the cavity Born-Oppenheimer (CBO)~\cite{flick2016} approximation that allows to construct approximate wave functions to the exact eigenstates for such problems. 
This work is structured into three sections: (i) First, the theoretical framework is introduced where we demonstrate how the concept of the Born-Oppenheimer approximation can be generalized to matter-photon coupled systems. (ii) We apply this theoretical framework to study a prototypical electron-photon system, where the photon couples resonantly to an electronic transition. (iii) The last section is devoted to a model system of a electron, a nuclei and photons, where a photon mode couples to a vibrational excitation.

\section{Theory}
\subsection{General correlated electron-nuclear-photon systems}
In what follows and without loss of generality, we describe the electron-nuclear-photon problem in Coulomb gauge, dipole approximation and the Power-Zienau-Woolley frame~\cite{babiker1983,faisal1987}. Our system of interest contains $n_e$ electrons, $n_n$ nuclei, and $n_p$ quantized photon modes, e.g. the matter is located in an optical high-Q cavity. Strong light-matter coupling is obtained, once the light-matter coupling is stronger than the dissipation of the system due to e.g. cavity losses. For simplicity, we neglect {dissipative channels in the following}. The original derivation of the Born-Oppenheimer approximation is outlined e.g. in Ref.~\cite{gross1991} for the specific case of electrons and ions and here we extend it to the photon case. In general, the correlated electron-nuclear-photon Hamiltonian can be written as follows~\cite{tokatly2013,pellegrini2015,flick2015,flick2016,craig1998}\footnote{Throughout this work, we assume SI units, unless stated otherwise.}. 
\begin{align}
\label{eq:correlated-hamiltonian}
\hat{H} = \hat{H}_e + \hat{H}_n + \hat{H}_{en} + \hat{H}_p +\hat{H}_{pe} +\hat{H}_{pn}+\hat{H}_{pen}
\end{align}
consisting of the electronic Hamiltonian $\hat{H}_e$ with $n_e$ electrons of mass $m_e$
\begin{align}
\label{eq:correlated-electronic-hamiltonian}
\hat{H}_e = \sum_{i=1}^{n_e} \frac{\hat{\textbf{p}}_i^2}{2m_e}+ \frac{e^2}{4\pi\epsilon_0}\sum_{i> j}^{n_e}\frac{1}{\left|\textbf{r}_i-\textbf{r}_j \right|},
\end{align}
the nuclear Hamiltonian $\hat{H}_n$ with $n_n$ nuclei each with possibly different individual masses $m_i$ and charges $Z_i$
\begin{align}
\label{eq:correlated-nuclear-hamiltonian}
\hat{H}_n = \sum_{i=1}^{n_n} \frac{\hat{\textbf{p}}_i^2}{2m_i}+ \frac{e^2}{4\pi\epsilon_0}\sum_{i> j}^{n_n}\frac{Z_iZ_j}{\left|\textbf{R}_i-\textbf{R}_j \right|} = \hat{T}_n + \hat{W}_n,
\end{align}
where $\hat{T}_n$ and $\hat{W}_n$ are the nuclear kinetic energy and nuclear {interaction}, respectively. The electron-nuclear interaction Hamiltonian $\hat{H}_{en}$ is given by
\begin{align}
\label{eq:correlated-electron-nuclear-hamiltonian}
\hat{H}_{en} = - \frac{e^2}{4\pi\epsilon_0}\sum_{i=1}^{n_e}\sum_{j=1}^{n_n}\frac{Z_j}{\left|\textbf{r}_i-\textbf{R}_j \right|},
\end{align}
and the cavity photon Hamiltonian $\hat{H}_p$ with $n_p$ quantized photon modes of frequency $\omega_\alpha$ {takes the form}
\begin{align}
\label{eq:correlated-photon-hamiltonian}
\hat{H}_p = \frac{1}{2}\left(\sum_{\alpha=1}^{2n_p}\hat{p}^2_\alpha + \omega_\alpha^2 \hat{q}_\alpha^2\right) = \hat{T}_p + \hat{W}_p.
\end{align}
The displacement field operators $\hat{q}_\alpha=\left(\hat{a}^\dagger_\alpha+\hat{a}_\alpha\right)/{\sqrt{2\omega_\alpha/\hbar}}$ consist of the usual photon creation and annihilation operators and $\left[\hat{q}_\alpha,\hat{p}_{\alpha'} \right]=i\hbar\delta_{\alpha,\alpha'}$. Furthermore, {the} $\hat{q}_\alpha$ are directly proportional to the electric displacement field operator of the $\alpha$-th photon mode~\cite{pellegrini2015,flick2015} at the charge-center of the system by the connection $\hat{\textbf{D}}_\alpha = \epsilon_0\omega_\alpha{\boldsymbol\lambda_\alpha}\hat{q}_\alpha$ {and the $\hat{p}_{\alpha}$ are proportional to the magnetic field. In Eq.~\ref{eq:correlated-photon-hamiltonian}, the sum runs from $1$ to $2n_p$, to correctly account for the two possible polarization directions of the electromagnetic field. The last three terms in Eq.~\ref{eq:correlated-hamiltonian} describe the light-matter interaction Hamiltonian. {The first term is} the explicit electron-photon interaction in the dipole approximation
\begin{align}
\label{eq:correlated-electron-photon-hamiltonian}
\hat{H}_{pe} = \sum_{\alpha=1}^{2n_p} \omega_\alpha \hat{q}_\alpha \left(\boldsymbol\lambda_\alpha \cdot \textbf{X}_e\right){,}
\end{align}
with the total electronic dipole moment $\textbf{X}_e = -\sum_{i=1}^{n_e} e\textbf{r}_i$ and the matter-photon coupling strength ${\boldsymbol \lambda_\alpha}$~\cite{tokatly2013,flick2015}. {The second term gives} the explicit nuclear-photon interaction{,} again in the dipole approximation{,}
\begin{align}
\label{eq:correlated-photon-nuclear-hamiltonian}
\hat{H}_{pn} = \sum_{\alpha=1}^{2n_p} \omega_\alpha \hat{q}_\alpha \left(\boldsymbol\lambda_\alpha \cdot \textbf{X}_n\right){,}
\end{align}
with the total nuclear dipole moment $\textbf{X}_n = \sum_{i=1}^{n_n} Z_ie\textbf{R}_i$. And {the last term describes} the quadratic dipole-self interaction term
\begin{align}
\label{eq:correlated-electron-nuclear-photon-hamiltonian}
\hat{H}_{pen} = \frac{1}{2}\sum_{\alpha=1}^{2n_p} \left(\boldsymbol\lambda_\alpha \cdot \textbf{X}\right)^2,
\end{align}
where $\textbf{X}$ now describes the total dipole moment of the system, i.e. $\textbf{X} = \textbf{X}_e + \textbf{X}_n$. We then introduce the following abbreviations
\begin{align*}
\underline{\textbf{r}} &= (\textbf{r}_1,...,\textbf{r}_{n_e})\\
\underline{\textbf{R}} &= (\textbf{R}_1,...,\textbf{R}_{n_N})\\
\underline{{q}} &= (q_{1},...,q_{2n_p}).
\end{align*}
Under this change of notation, we can rewrite Eq.~\ref{eq:correlated-hamiltonian} in the following form
\begin{align}
\label{eq:correlated-hamiltonian-2}
\hat{H} &= \hat{H}(\underline{\textbf{r}},\underline{\textbf{R}},\underline{q}) =  \hat{H}_e(\underline{\textbf{r}}) + \hat{H}_n(\underline{\textbf{R}}) + \hat{H}_{en}(\underline{\textbf{r}},\underline{\textbf{R}}) \nonumber\\
& + \hat{H}_p(\underline{q})+ \hat{H}_{pe}(\underline{\textbf{r}},\underline{q}) +\hat{H}_{pn}(\underline{\textbf{R}},\underline{q})+\hat{H}_{pen}(\underline{\textbf{r}},\underline{\textbf{R}}).
\end{align}
In general, we are interested in calculating eigenstates $\Psi_i(\underline{\textbf{r}},\underline{\textbf{R}},\underline{q})$ and eigenvalues $E_i$ of the particular problem. These states then give us access to any observable of interest. To calculate these quantities, we have to solve the full Schr\"odinger equation of the correlated electron-nuclear-photon problem that is given by
\begin{align}
\label{eq:correlated-schroed}
\hat{H} \Psi_i(\underline{\textbf{r}},\underline{\textbf{R}},\underline{q}) = E_i \Psi_i(\underline{\textbf{r}},\underline{\textbf{R}},\underline{q}),
\end{align}
where the Hamiltonian $\hat{H}$ is given by Eq.~\ref{eq:correlated-hamiltonian}. Obtaining general solutions to the Schr\"odinger equation of Eq.~\ref{eq:correlated-schroed} is an ungrateful task\footnote{We note that in free space Eq.~\ref{eq:correlated-schroed} has no eigenstates due to its translational invariance. Hence one either has to go into a co-moving frame, e.g., a center-of-mass frame, or one has to use a confining potential to localize the molecule.}. In practice, the Schr\"odinger equation is barely solved exactly, but only approximately. One of such approximate methods is the cavity Born-Oppenheimer approximation~\cite{flick2016} that is capable to partially decouple the electronic degrees of freedom from the nuclear and photonic degrees of freedom. In electron-nuclear problems, such an adiabatic decoupling procedure is commonly assumed~\cite{gross1991} and well justified for low lying states, i.e. the ground state. However, severe limitations are known that require going beyond the adiabatic treatment by including nonadiabatic electron-nuclear terms, e.g. at conical intersections~\cite{li2010}. 
\\
In this work we decouple the electronic degrees of freedom from the nuclear and photon degrees of freedom. This allows us, on the one hand, to simplify the problem much more than if we decoupled the nuclear from the electronic and photonic degrees of freedom, as has been done in Refs.~\cite{galego2015,galego2016}. In practice, the main problem for the standard Born-Oppenheimer approximation is to solve the resulting electronic equation, while simple approximations to the nuclear equation, such as harmonic approximations, are often sufficient. On the other hand, a decoupling of the electronic degrees of freedom provides most flexibility for the applications that we consider, e.g. a single electron coupled to one mode. From a physical perspective, however, this decoupling scheme seems counterintuitive on a first glance. The usual simplified argument for the decoupling of the nuclear from the electronic degrees of freedom is that the nuclei move ``slowly'' compared to the electrons, i.e., the kinetic-energy contribution is negligible, and hence a classical approximation seems reasonable. Photons do not move ``slowly'' and hence a similar simple argument does not make much sense. However, for the photons the term $\hat{T}_p$ describes the square of the magnetic field operator and is a small perturbation compared to the harmonic potential that confines the mode. In this sense, a classical approximation for the photons is reasonable and physically means that we neglect the magnetic contribution to the photon-field energy. That this approach can indeed give highly accurate results will be demonstrated in the following.

\subsection{Cavity Born-Oppenheimer approximation}

In this section, we derive the approximate cavity Born-Oppenheimer states to Eq.~\ref{eq:correlated-schroed}. This goal is achieved in three successive steps. First, we solve the electronic part of the Eq.~\ref{eq:correlated-schroed}, where we consider explicitly all terms containing an explicit electronic contribution. This electronic Schr\"odinger equation has only a parametric (conditional) dependence on the nuclear and field degrees of freedom, or alternatively nuclear and field {coordinates} enter the electronic equation as {c-numbers}. In principle, the electronic Schr\"odinger {equation} has to be solved for every possible {combined} nuclear {and photon-}field configuration and the eigenvalues of the electronic Schr\"odinger equation then enter the nuclear {and photon-field} Schr\"odinger equation through the emerging potential-energy surfaces. Having solved both equations, we can then construct the approximate cavity Born-Oppenheimer states in a factorized manner. To obtain the approximate cavity Born-Oppenheimer states, as a first step, we solve the electronic Schr\"odinger equation
\begin{align}
\label{eq:electron-schrodinger}
&\left[\hat{H}_e(\underline{\textbf{r}}) + \hat{H}_{en}(\underline{\textbf{r}},\underline{\textbf{R}}) + \hat{H}_{pe}(\underline{\textbf{r}},\underline{q})+ \hat{H}_{pen}(\underline{\textbf{r}},\underline{\textbf{R}}) \right]\psi_j(\underline{\textbf{r}},\underline{\textbf{R}},\underline{q}) \nonumber\\
&= \epsilon_j(\underline{\textbf{R}},\underline{q})\psi_j(\underline{\textbf{r}},\underline{\textbf{R}},\underline{q}),
\end{align}
for each fixed set of nuclear coordinates $\underline{\textbf{R}}$, and photon displacement coordinates $\underline{q}$. For each fixed set of $(\underline{\textbf{R}},\underline{q})$, the electronic eigenfunctions of Eq.~\ref{eq:electron-schrodinger} $\left\{\psi_j(\underline{\textbf{r}},\underline{\textbf{R}},\underline{q})\right\}$ form a complete basis in the electron many-particle Hilbert space. In the electronic Schr\"odinger equation of Eq.~\ref{eq:electron-schrodinger}, $(\underline{\textbf{R}},\underline{q})$ enter the electronic cavity Born-Oppenheimer Hamiltonian as (classical) parameters, thus the eigenvalues $\epsilon_j$ also parametrically depend on $\underline{\textbf{R}},\underline{q}$. For each fixed set of $(\underline{\textbf{R}},\underline{q})$, we can then expand (also known as the Born-Huang expansion~\cite{born1956}) the exact many-body wave function $\Psi_i(\underline{\textbf{r}},\underline{\textbf{R}},\underline{q})$ that is a solution to the full Schr\"odinger equation of Eq.~\ref{eq:correlated-schroed} as 
\begin{align}
\Psi_i(\underline{\textbf{r}},\underline{\textbf{R}},\underline{q}) = \sum_{j=1}^{\infty}\chi_{ij}(\underline{\textbf{R}},\underline{q})\psi_j(\underline{\textbf{r}},\underline{\textbf{R}},\underline{q}).
\end{align}
Here, the exact wave function is decomposed into sums of product states consisting of an electronic wave function $\psi_j(\underline{\textbf{r}},\underline{\textbf{R}},\underline{q})$ and a nuclear-photon wave function  $\chi_{ij}(\underline{\textbf{R}},\underline{q})$. The latter is obtained by solving the following equation
\begin{align}
&\left[\hat{H}_n(\underline{\textbf{R}}) + \hat{H}_p(\underline{q}) +\hat{H}_{pn}(\underline{\textbf{R}},\underline{q}) + \epsilon_{k}(\underline{\textbf{R}},\underline{q})\right] \chi_{ik}(\underline{\textbf{R}},\underline{q}) \nonumber\\
+&\sum\limits_{j=1}^{\infty} \left(\int d\underline{\textbf{r}} \psi_k^*(\underline{\textbf{r}},\underline{\textbf{R}},\underline{q}) \left[\hat{T}_n(\underline{\textbf{R}}) + \hat{T}_p(\underline{q})\right]\psi_j(\underline{\textbf{r}},\underline{\textbf{R}},\underline{q})\right)\chi_{ij}(\underline{\textbf{R}},\underline{q})\nonumber\\
&=E_i \chi_{ik}(\underline{\textbf{R}},\underline{q}),
\label{eq:fp-photon-hamiltonian}
\end{align}
where $\hat{T}_n(\underline{\textbf{R}})$ and $\hat{T}_p(\underline{q})$ are given by Eqns.~\ref{eq:correlated-nuclear-hamiltonian} and~\ref{eq:correlated-photon-hamiltonian}, respectively. The eigenvalues $E_i$ of Eq.~\ref{eq:fp-photon-hamiltonian} are the exact correlated eigenvalues of Eq.~\ref{eq:correlated-schroed}. The term in the second line of Eq.~\ref{eq:fp-photon-hamiltonian} describes the nonadiabatic coupling between cavity Born-Oppenheimer potential energy surfaces (PES). The cavity Born-Oppenheimer approximation now neglects the offdiagonal elements in the nonadiabatic coupling terms of Eq.~\ref{eq:fp-photon-hamiltonian}. Then Eq.~\ref{eq:fp-photon-hamiltonian} can be rewritten in a much simpler form
\begin{align}
&\left[ \hat{T}_n(\underline{\textbf{R}}) + \hat{T}_p(\underline{q}) + V_k(\underline{\textbf{R}},\underline{q})\right] \chi_{ik}(\underline{\textbf{R}},\underline{q}) \nonumber\\
&=E_i \chi_{ik}(\underline{\textbf{R}},\underline{q}),
\label{eq:fp-photon-hamiltonian-boa}
\end{align}
where the newly generalized {cavity PES} $V_j(\underline{\textbf{R}},\underline{q})$ are given explicitly by
\begin{align}
\label{eq:CBOA-surfaces}
V_j(\underline{\textbf{R}},\underline{q}) &=\hat{W}_n(\underline{\textbf{R}}) + \hat{W}_p(\underline{q}) + \hat{H}_{pn}(\underline{\textbf{R}},\underline{q}) \nonumber\\
&+\int d\underline{\textbf{r}} \ \psi_j^*(\underline{\textbf{r}},\underline{\textbf{R}},\underline{q}) \left[\hat{T}_n(\underline{\textbf{R}}) + \hat{T}_p(\underline{q})\right]\psi_j(\underline{\textbf{r}},\underline{\textbf{R}},\underline{q})\nonumber\\
&+ \epsilon_j(\underline{\textbf{R}},\underline{q}).
\end{align}
The first two terms are the nuclear and the photon potentials of Eqns.~\ref{eq:correlated-nuclear-hamiltonian} and~\ref{eq:correlated-photon-hamiltonian} and all anharmonicity in the PES can be attributed to the electron-photon, electron-nuclear, nuclear-nuclear and nuclear-photon interaction contained in Eq.~\ref{eq:correlated-hamiltonian}. Furthermore, the eigenvalues $E_i$ of Eq.~\ref{eq:fp-photon-hamiltonian-boa} are an approximation to the exact correlated eigenvalues and provide by the variational principle an upper bound. With this reformulation, we have the advantage that we can solve the electronic Schr\"odinger equation of Eq.~\ref{eq:electron-schrodinger} and the nuclear-photon Schr\"odinger Eq.~\ref{eq:fp-photon-hamiltonian-boa} separately. {The ground-state $\Psi_0$ in the cavity Born-Oppenheimer approximation then becomes}
\begin{align}
 \Psi_{0,CBO}(\underline{\textbf{r}},\underline{\textbf{R}},\underline{q})  = \chi_{00}(\underline{\textbf{R}},\underline{q})\psi_0(\underline{\textbf{r}},\underline{\textbf{R}},\underline{q}),
\end{align}
{and accordingly for the excited states.} In Born-Oppenheimer calculations {for systems that only contain electrons and} nuclei often the harmonic Born-Oppenheimer approximation is carried out~\cite{gross1991} that can be realized by expanding $V_j(\underline{\textbf{R}},\underline{q})$ around its minimum value and {in this way} even simplifies the problem further. In the harmonic approximation, we have to solve Eq.~\ref{eq:electron-schrodinger} not for all possible values of $(\underline{\textbf{R}},\underline{q})$, but only at the minimum of $\epsilon_j(\underline{\textbf{R}},\underline{q})$. However, in this work, we do not apply the harmonic approximation to correctly demonstrate the full capacity of the cavity Born-Oppenheimer concept.\\
Before we introduce our the examples, let us comment on the expectable accuracy of the cavity Born-Oppenheimer states when decoupling electronic from photonic and nuclear degrees of freedom. Our simplified physical arguments for the decoupling scheme so far have been that the nuclei are ``slow'' and the magnetic-field contribution small, such that we can neglect the corresponding kinetic terms in the equation for the electronic subsystem. {However, the decisive quantities that indicate the quality of this approach are the nonadiabatic coupling elements of Eq.~\ref{eq:fp-photon-hamiltonian} and the distance between the the potential-energy surfaces. If these elements are small and the potential-energy surfaces are far apart, we can expect a good quality of the approximate cavity Born-Oppenheimer states. This argument is similar to standard Born-Oppenheimer treatment that looses its validity at crossing of eigenvalues, i.e. conical intersections.}

\section{Discussion and Results}

In the following, we now want to illustrate the concept of the cavity Born-Oppenheimer {approximation} for two specific setups. We numerically analyze first a model system consisting of a single electron coupled resonantly to a photon mode. This model will allow us to study the decoupling mechanism introduced for the correlated electron-photon interaction in detail. In the second example, we then analyze a model system that contains electron-nuclear-field degrees of freedom. Here, potential-energy surfaces emerge that have nuclear-photon (polaritonic) nature.

\subsection{Light-Matter coupling via electronic excitation}
\label{sec:gaas}

\begin{figure}[ht] 
\centerline{\includegraphics[width=0.5\textwidth]{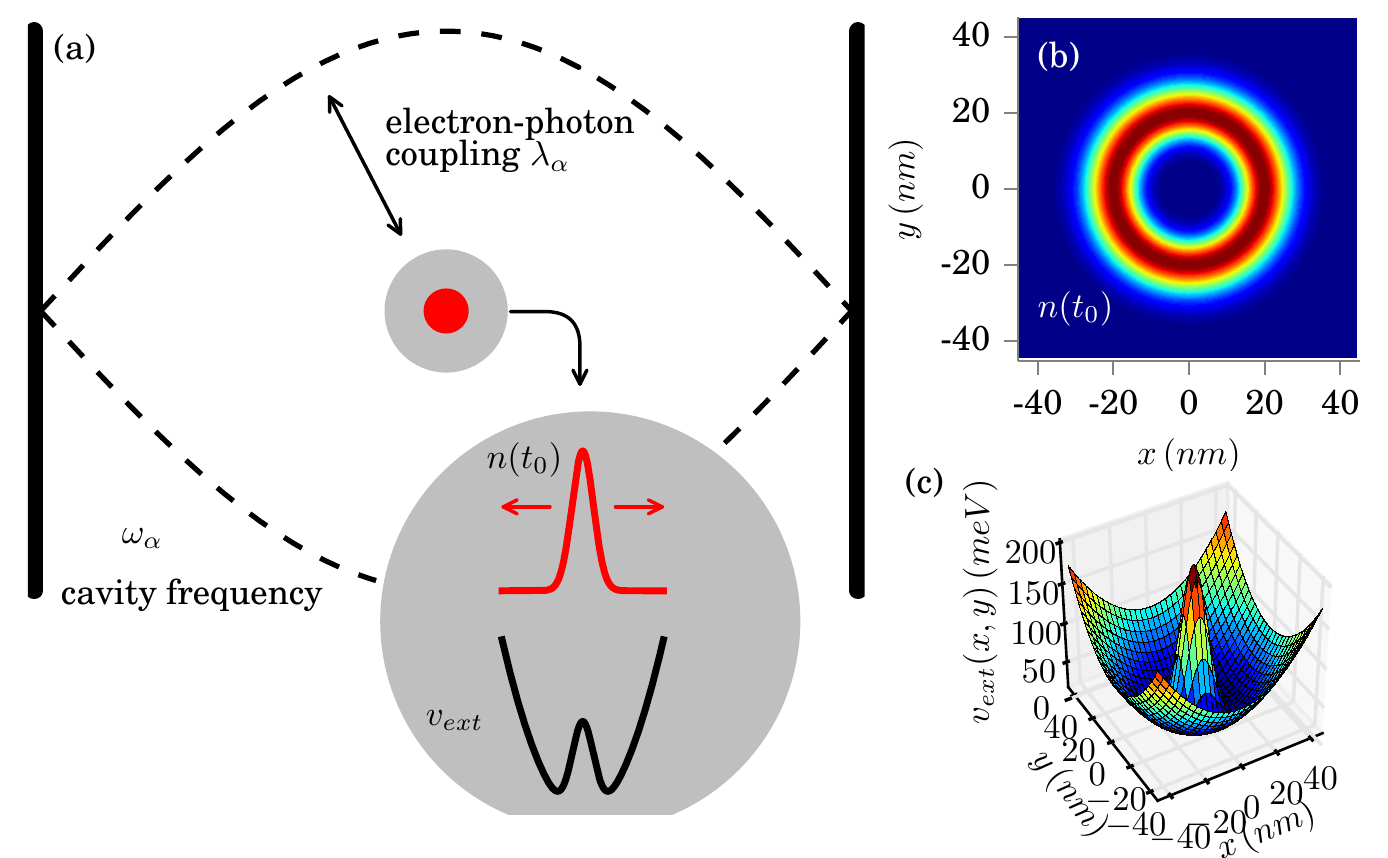}}
\caption{(a) Model for the GaAs quantum ring in an optical cavity. (b) Bare ground-state electron density $n_{\lambda=0}$ in the external potential that is shown in (c).}
\label{fig:gaas-model}
\end{figure}
In this section, we illustrate the concept of the cavity Born-Oppenheimer {approximation} for a simple coupled electron-photon model system. The system of interest is a model system for a GaAs quantum ring~\cite{rasanen2007} that is located in an optical cavity and thus coupled to a single photon mode~\cite{flick2015}. The model features a single electron confined in two-dimensions in real-space ($\textbf{r}=r_x\textbf{e}_x + r_y\textbf{e}_y$) interacting with the single photon mode with frequency $\hbar\omega_\alpha=1.41$~meV and polarization direction $\textbf{e}_\alpha=(1,1)$. The polarization direction enters via the electron-photon coupling strength, i.e. ${\boldsymbol \lambda_\alpha}=\lambda_\alpha \textbf{e}_\alpha$. The photon mode frequency is chosen to be in resonance with the first electronic transition. We depict the model schematically in Fig.~\ref{fig:gaas-model} (a).
The bare electron ground-state $n_{\lambda=0}(\textbf{r})$ has a ring-like structure shown in Fig.~\ref{fig:gaas-model} (b) due to the Mexican-hat like external potential that is given by
\begin{align}
v_{ext}(\textbf{r}) = \frac{1}{2}m_0 \omega_0^2 \textbf{r}^2 + V_0 e^{-\textbf{r}^2/d}{,}
\end{align}
with parameters $\hbar\omega_0=10$~meV, $V_0=200$~meV, $d=10$~nm~\cite{rasanen2007}, and shown in Fig.~\ref{fig:gaas-model} (c). For the single electron, we employ a two-dimensional grid of $N=127$ grid points in each direction with $\Delta x=0.7052$~nm. In contrast, we include the photons for the exact calculation in the photon number eigenbasis, where we include up to $41$ photons in the photon mode.\\ 
For the cavity Born-Oppenheimer calculations, we calculate the photons also on an uniform real-space grid (q-representation) with $N=41$ with $\Delta q= 6.77$~$\sqrt{\text{aJ}}$~fs$^2$ and construct the projector from the uniform real-space grid to the photon number states basis explicitly. This projector can be calculated by employing the eigenstates of the quantum harmonic oscillator in real-space. For a more detailed discussion of the model system, we refer the reader to Refs.~\cite{rasanen2007,flick2015}. Since this model can be solved by exact diagonalization in full Fock space~\cite{flick2014}, all exact results shown in the following have been calculated employing the full correlated electron-photon Hamiltonian~\cite{tokatly2013,pellegrini2015,flick2015,flick2016}.
\begin{figure}[ht] 
\centerline{\includegraphics[width=0.5\textwidth]{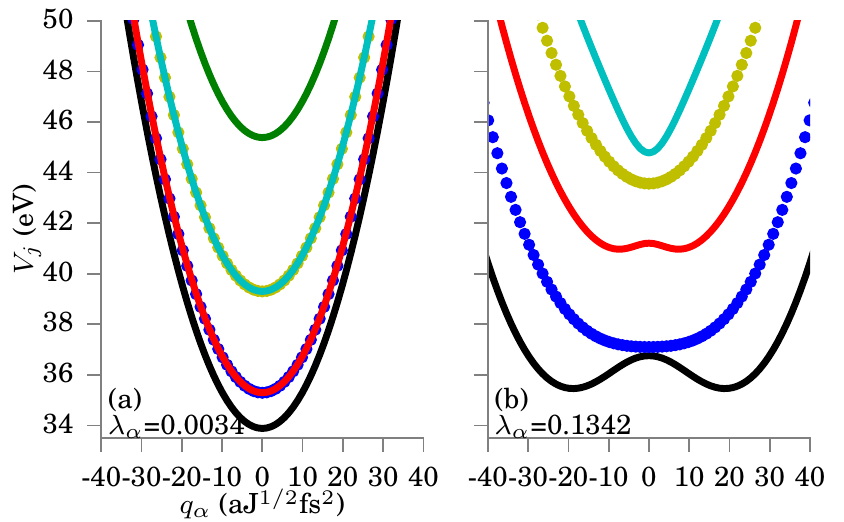}}
\caption{Born-Oppenheimer potential energy surfaces $V_j$ for a correlated electron-photon problem in (a) weak coupling with $\lambda_\alpha=0.0034$  meV$^{1/2}$/nm and (b) strong coupling $\lambda_\alpha=0.1342$  meV$^{1/2}$/nm.}
\label{fig:qedft-fp-bo-surfaces}
\end{figure}
For this model, the potential-energy surfaces from Eq.~\ref{eq:CBOA-surfaces} can be calculated explicitly as
\begin{align}
\label{eq:CBOA-surfaces-dot}
V_j({q_\alpha}) &= \frac{1}{2}\omega_\alpha q^2_\alpha + \epsilon_j({q_\alpha})\nonumber\\
&+\int d{\textbf{r}} \ \psi_j^*({\textbf{r}},{q_\alpha}) \hat{T}_p({q_\alpha})\psi_j({\textbf{r}},{q_\alpha}).
\end{align}
In Fig.~\ref{fig:qedft-fp-bo-surfaces} (a), we show the PES surfaces $V_j(\left\{q_\alpha\right\})$ for the weak-coupling regime of $\lambda_\alpha = 0.0034$~meV$^{1/2}$/nm. We find that all {PES} have a strong harmonic nature, due to the dominant $\hat{q}_{\alpha}^2$ term in Eq.~\ref{eq:CBOA-surfaces-dot}. {The eigenvalues} $\epsilon_j$ and the integral in the last line of Eq.~\ref{eq:CBOA-surfaces-dot} are the corrections to the harmonic potential. In this case, both are rather small for all excited-state surfaces in the weak-coupling regime, i.e. for the ground-state surface adiabatic term in the last line of Eq.~\ref{eq:CBOA-surfaces-dot} is around two orders of magnitude smaller than  $\epsilon_0$. In general, a harmonic correction that can be obtained by calculating the second derivative at the minimum value will shift the frequency of the photon mode. We define as harmonic approximation to Eq.~\ref{eq:CBOA-surfaces-dot}
\begin{align}
\label{eq:CBOA-surfaces-dot-harmonic}
V_{j,harm}({q_\alpha}) &= \frac{1}{2}\tilde{\omega}_{j,\alpha} \left(q_\alpha-q_{j,0}\right)^2,
\end{align}
where $q_{j,0}$ is the minimum value of the $j$-th PES Eq.~\ref{eq:CBOA-surfaces-dot}. In the weak-coupling regime, we find $\tilde{\omega}_\alpha\approx {\omega}_{j,\alpha}$. All corrections beyond the second derivative of these terms are then called the anharmonic corrections.\\
We find the lowest {cavity} {PES} that is the ground-state PES shown in black, well separated from the first and second excited {cavity} {PES} that are  shown in solid red and dotted blue. The first and second excited {cavity} {PES} are close to being degenerate. This two-fold degeneracy has its origin in the two-dimensional external potential, similar to the $s$/$p$ degeneracy in the hydrogen atom. In Fig.~\ref{fig:qedft-fp-bo-surfaces} (b), we show the {cavity} {PES} surfaces in the strong-coupling regime with $\lambda_\alpha = 0.134$~meV$^{1/2}$/nm. While the second PES shown in blue and the fourth potential energy surface shown in yellow keep the harmonic shape, in the lowest {cavity} {PES} shown in black and the third {cavity} {PES} shown in solid red, two new minima with a double-well structure appear \footnote{{Note that if we would like to express this electron-dressed photon system in terms of the original creation and annihilation operators, we will need new combinations of these operators, i.e., photon-interaction terms. Physically these interaction terms describe the coupling between photons mediated via the electron.}}. The minima of the {cavity} {PES} are strongly shifted  away from the equilibrium position at the origin. This electron-dressed potential for the photon modes induces a new vacuum state with two maxima. {Since the cavity PES is symmetric, the vacuum state still has a displacement observable of $\langle q_\alpha\rangle=0$, i.e., we have a stable vacuum with zero field. However, with respect to the bare vacuum the other observables, e.g., the vacuum fluctuations, will clearly change.} Furthermore, we find for the harmonic approximation in the ground-state {cavity} PES, $\tilde{\omega}_{0,\alpha}\approx0.8 {\omega}_{\alpha}$, hence an effective softening of the photon mode in the ground-state {cavity} PES with the strong displacement of $q_{0,0}=18.85$~$\sqrt{\text{aJ}}$~fs$^2$. A similar behavior has been observed before in the context of polaron physics in the Holstein Hamiltonian~\cite{sakkinen2014,sakkinen2015}.
\begin{figure}[ht] 
\centerline{\includegraphics[width=0.5\textwidth]{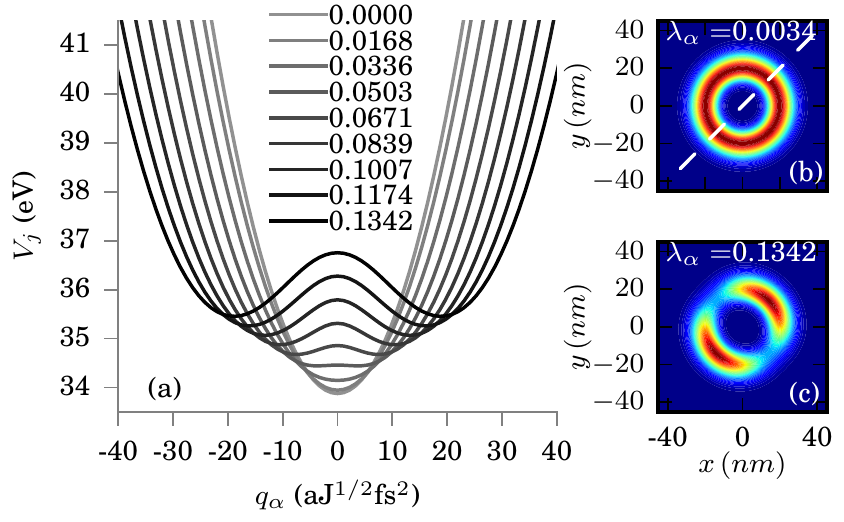}}
\caption{Left: (a) Ground-state {cavity} PES for different coupling strengths show an emerging displacement of the photon states. Right: electron density in (a) the weak coupling regime for $\lambda_\alpha=0.0034$~meV$^{1/2}$/nm and (b) strong coupling for  $\lambda_\alpha=0.1342$~meV$^{1/2}$/nm. The dashed-lines in (b) indicate the polarization direction $\textbf{e}_\alpha$ of the photon mode and red color refers to high-density regions, while blue color refers to low-density regions.} 
\label{fig:qedft-fp-bo-surfaces-ground-state}
\end{figure}
We further analyze this transition in Fig.~\ref{fig:qedft-fp-bo-surfaces-ground-state}. In Fig.~\ref{fig:qedft-fp-bo-surfaces-ground-state} (a), we show how the ground-state PES depends on the electron-photon coupling strength ${ \lambda_\alpha}$. We find that for absent and weak coupling, the ground-state surface can be well described by a single harmonic potential that has the minimum at $q_\alpha=0$. If we increase the electron-photon coupling to strong coupling, we find around $\lambda_\alpha=0.044$~meV$^{1/2}$/nm the splitting of the single{-}well structure to a double{-}well structure. For strong coupling, e.g. $\lambda_\alpha=0.1342$~meV$^{1/2}$/nm this double{-}well structure becomes strongly pronounced. In Fig.~\ref{fig:qedft-fp-bo-surfaces-ground-state} (b) and (c), we plot the corresponding electron density {$n_\lambda(\textbf{r})=\int dq_\alpha\Psi^*_{0,\lambda}(\textbf{r},q_\alpha)\Psi_{0,\lambda}(\textbf{r},q_\alpha)$} of the exact correlated ground state {$\Psi_{0,\lambda}(\textbf{r},q_\alpha)$ for different values of $\lambda$}. In the weak-coupling regime, shown in Fig.~\ref{fig:qedft-fp-bo-surfaces-ground-state} (b), we find that the electron is only slightly distorted in comparison to the ring-like structure of the bare electron ground state~\cite{flick2015} shown in Fig.~\ref{fig:gaas-model} (b). In contrast, in the strong coupling regime, shown in Fig.~\ref{fig:qedft-fp-bo-surfaces-ground-state} (c), the electron density becomes spatially separated and localized in direction of the polarization direction of the quantized photon mode.\\
The consequences of the ground-state transition identified in Fig.~\ref{fig:qedft-fp-bo-surfaces-ground-state} become also apparent if we study the difference of the correlated and bare electron density. Let us define the bare electron density. Here, we refer to the electron density that is the ground-state of the external potential {without coupling to the photon mode}, or alternatively $\lambda_\alpha=0$, thus $n_{\lambda=0}(\textbf{r})$. This density is shown in Fig.~\ref{fig:gaas-model} (b). Then we define {$\Delta n_\lambda(\textbf{r})=n_\lambda(\textbf{r})-n_{\lambda=0}(\textbf{r})$}.
\begin{figure}[ht] 
\centerline{\includegraphics[width=0.5\textwidth]{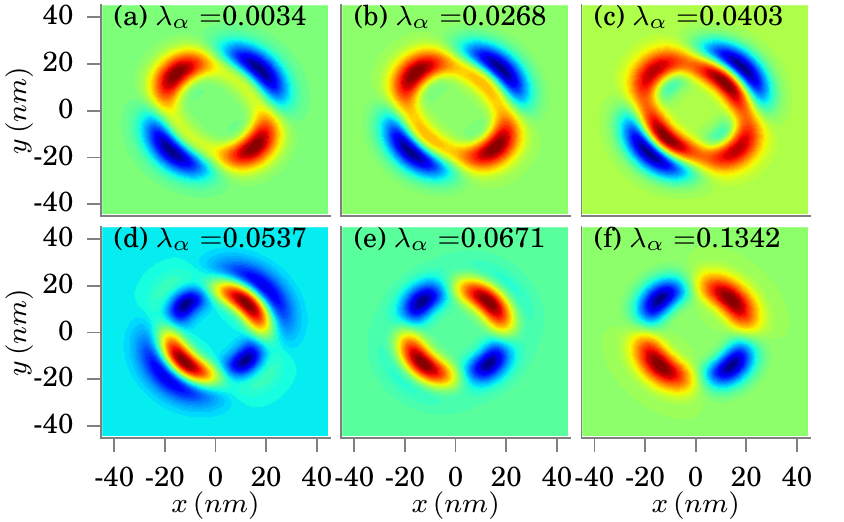}}
\caption{The difference of the correlated ground-state electron density to the bare electron density ({$\Delta n_\lambda=n_\lambda-n_{\lambda=0}$}) from the weak- to the strong-coupling limit. Red color refers to surplus density regions, while blue color refers to {regions with reduced density.}}
\label{fig:density-transition}
\end{figure}
In Fig.~\ref{fig:density-transition}, we plot $\Delta n_\lambda(\textbf{r})$ as function of the electron-photon coupling strength $\lambda_\alpha$. In the weak-coupling limit, shown in Fig.~\ref{fig:density-transition} (a) for $\lambda_\alpha=0.0034$~meV$^{1/2}$/nm, we find that the electron density is slightly distorted such that in the correlated density more density is accumulated perpendicular to the polarization direction of the photon mode {compared to} the bare electron density. However, once the strong-coupling regime is approached, we also identify a transition in $\Delta n_\lambda(\textbf{r})$. In the strong coupling regime, that is entered in  Fig.~\ref{fig:density-transition} (b)-(d), the ground-state electron density is reoriented until ultimately in Fig.~\ref{fig:density-transition} (e) the electron density is arranged in direction of the polarization direction of the photon mode, up to higher strong-coupling regions shown in Fig.~\ref{fig:density-transition} (f).\\
The additional insights from the ground-state transition can be obtained by evaluating the exact correlated electron-photon eigenvalues.
\begin{figure}[ht] 
\centerline{\includegraphics[width=0.5\textwidth]{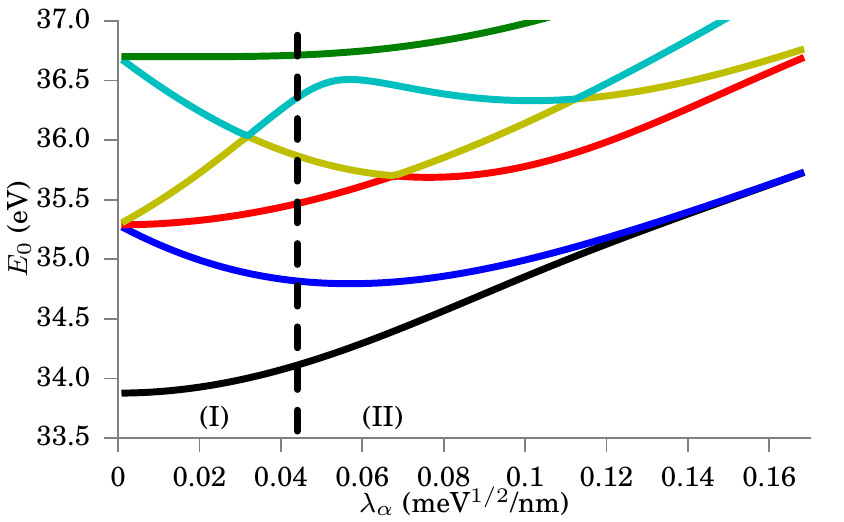}}
\caption{Exact eigenvalues of the correlated electron-photon Hamiltonian as function of the electron-photon coupling parameter $\lambda_\alpha$. The dashed line indicates the transition of $\Delta n_\lambda(\textbf{r})$ as discussed in the main text.}
\label{fig:ground-state-energies}
\end{figure}
In Fig.~\ref{fig:ground-state-energies}, we plot the exact eigenvalues from the weak- to the strong-coupling regime. The ground-state energies are plotted by the black line and are increasing for stronger coupling~\cite{flick2016}. For the first excited state in the case of $\lambda_\alpha=0$ coupling, we find a three-fold degeneracy that is split once the electron-photon coupling is introduced. For strong coupling the first-excited state (shown in blue) and the ground-state become close\footnote{We emphasize that this behavior is similar to what is in molecular systems known as \textit{static correlation} for e.g. stretched molecules~\cite{dimitrov2016}.} leading to the splitting of the electron-density shown in Fig.~\ref{fig:density-transition}.
Higher-lying states show energy crossings that are typical for electron-photon problems and have been previously observed e.g. in the Rabi model~\cite{braak2011,xie2016,boite2016}. We find allowed level crossings at $\lambda_\alpha\approx 0.031,0.067,0.113$~meV$^{1/2}$/nm, but also an {avoided} level crossing at $\lambda_\alpha\approx 0.55$~meV$^{1/2}$/nm between the fifth and sixth eigenvalue surface. In the Rabi model, level crossings are used to define transition from the weak, strong, ultra-strong~\cite{bamba2015} and deep-strong coupling regime~\cite{casanova2010}. Similarly to the Rabi model~\cite{braak2011}, we find in the strong coupling regime a pairing of states in terms of the energy. Two states each with different parity become close to degeneracy. Since in the strong-coupling regime the interaction terms in the Hamiltonian become dominant and we apply the interaction in dipole coupling, the eigenstates of the full Hamiltonian become close to the eigenstates of the dipole operator that are the parity eigenstates. We can expect a different behavior beyond the dipole coupling, e.g. if electric quadrupole and magnetic dipole coupling{, or higher multipolar coupling terms are} also considered. In Fig.~\ref{fig:ground-state-energies}, we indicate by the dashed line, the ground-state transition discussed before. In the coupling region indicated by (I), we find a single minimum in the PES and $\Delta n$ is located perpendicular to the polarization direction, while in the coupling regime (II), we find two minima and a double well structure in the PES and $\Delta n$ is located along the direction of the polarization  of the photon mode.%
\begin{center}%
\begin{table}%
\renewcommand{\tabcolsep}{2.5mm}%
\begin{tabular}{| c| c| c | c | c |c |}%
\hline%
state \# & $\lambda_\alpha$ & $E^{\text{exact}}$ & {$E_{CBO}$} &  (e,n)    & overlap\\\hline\hline%
1 & 0.0034 & 33.8782 & 33.8795 & 1,1 & 99.9539\\%
2 & 0.0034 & 35.2293 & 35.2861 & 1,2 & 55.7957\\%
3 & 0.0034 & 35.2898 & 35.2898 & 2,1 & 99.9992\\%
4 & 0.0034 & 35.3521 & 35.2979 & 3,1 & 55.8438\\%
5 & 0.0034 & 36.6153 & 36.6925 & 1,3 & 57.4860\\%
\hline%
\hline%
1 & 0.0302 & 33.9902 & 34.0258 & 1,1 & 98.7922\\%
2 & 0.0302 & 34.8957 & 35.0935 & 1,2 & 84.9288\\%
3 & 0.0302 & 35.3734 & 35.3763 & 2,1 & 99.9475\\%
4 & 0.0302 & 35.9902 & 35.8670 & 3,1 & 84.4187\\%
5 & 0.0302 & 36.0575 & 36.2793 & 1,3 & 86.7428\\%
\hline%
\hline%
1 & 0.0637 & 34.3433 & 34.3659 & 1,1 & 99.3180\\%
2 & 0.0637 & 34.8006 & 34.9008 & 1,2 & 96.1220\\%
3 & 0.0637 & 35.6546 & 35.6613 & 2,1 & 99.8841\\%
4 & 0.0637 & 35.7142 & 35.8487 & 1,3 & 94.9875\\%
5 & 0.0637 & 36.4857 & 36.7584 & 1,4 & 79.8066\\%
\hline%
\hline%
1 & 0.1342 & 35.3072 & 35.3114 & 1,1 & 99.9413\\%
2 & 0.1342 & 35.3307 & 35.3398 & 1,2 & 99.8537\\%
3 & 0.1342 & 36.1782 & 36.1953 & 1,3 & 99.6475\\%
4 & 0.1342 & 36.4492 & 36.4860 & 1,4 & 99.2544\\%
5 & 0.1342 & 36.7302 & 36.7345 & 2,1 & 99.9373\\%
\hline%
\hline%
\end{tabular}%
\caption{Exact correlated energies $E^{\text{exact}}$ (eV), {cavity} BO energies {$E_{CBO}$} (eV) and overlap between exact and {cavity} BO states depending on the electron-photon coupling strength $\lambda_\alpha$ given in meV$^{1/2}$/nm.  The label (e,n)  refers to the {cavity} BO quantum number of the state/excitation (electronic state, photon state). Note that {we do not employ the harmonic approximation and that} the {cavity} BO energies {$E_{CBO}$} provide an upper bound to the exact correlated energies $E^{\text{exact}}$.}%
\label{tab:overlap}%
\end{table}%
\end{center}%
The quality of the cavity Born-Oppenheimer  {approximation} is shown in Tab.~\ref{tab:overlap} in terms of overlaps $\langle\Psi_j| \Psi_{j,CBO}\rangle^2$ between approximate and exact states. If the eigenenergies shown in Fig.~\ref{fig:ground-state-energies}, are well separated as in the strong coupling regime for $\lambda_\alpha=0.1342$~meV$^{1/2}$/nm, {then} the cavity Born-Oppenheimer {approximation} is well justified. For states that are close to degeneracy, as e.g. the states \#2 and \#4 in the weak-coupling for $\lambda_\alpha=0.0034$~meV$^{1/2}$/nm, we find a lower quality. However, this low quality could be improved by symmetry considerations. Overall, we find a very high and sufficient quality of the approximate energies and states in comparison to its corresponding exact values.\\
\\
The remaining part of this section is concerned with the time-dependent case. Here, we employ the full correlated electron-photon Hamiltonian and choose as initial state a factorized initial state that consists of the bare electronic ground state and {a bare} photon field in a coherent state with $\langle \hat{a}^\dagger\hat{a}\rangle=4$ where $\lambda_\alpha=0.0034$~meV$^{1/2}$/nm. This example is also the first time-dependent example studied in  Ref.~\cite{flick2015}. To numerically propagate the system, we use a Lanczos scheme and propagate the initial state in 160000 time steps with {$\Delta t = 0.146$~fs}.
\begin{figure}[h] 
\centerline{\includegraphics[width=0.5\textwidth]{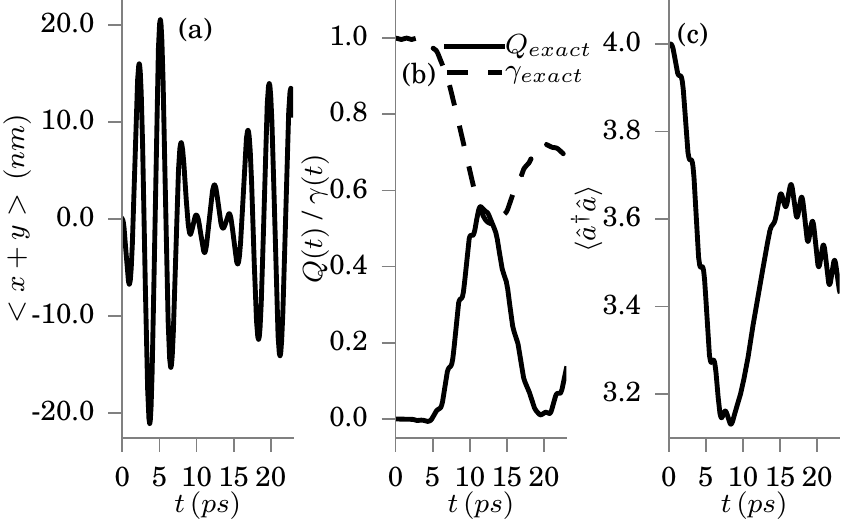}}
\caption{Time-dependent calculation with a factorizable initial state (a) dipole moment of the system, (b) Mandel $Q$ parameter and purity $\gamma$ and (c) the photon occupation $\langle \hat{a}^\dagger\hat{a}\rangle$ evolving in time.}
\label{fig:td-dipole-moment}
\end{figure}
In Fig.~\ref{fig:td-dipole-moment}, we briefly analyze this setup by evaluating the dipole moment $\langle \hat{x} + \hat{y}\rangle$ in Fig.~\ref{fig:td-dipole-moment} (a), the purity $\gamma = \text{Tr}\left(\rho_{ph}^2 \right)$ that contains the reduced photon density matrix $\rho_{ph}$ and the Mandel $Q$ parameter~\cite{mandel1979} that is defined as
\begin{align}   
Q = \frac{\langle \hat{a}^\dagger_{\alpha} \hat{a}^\dagger_{\alpha} \hat{a}_{\alpha} \hat{a}_{\alpha} \rangle -  \langle \hat{a}^\dagger_{\alpha} \hat{a}_{\alpha} \rangle^2 }{\langle \hat{a}^\dagger_{\alpha} \hat{a}_{\alpha} \rangle},
\end{align} 
in Fig.~\ref{fig:td-dipole-moment} (b) and the photon occupation $\langle \hat{a}^\dagger\hat{a}\rangle$ in Fig.~\ref{fig:td-dipole-moment} (c). In the case of the dipole moment of this example shown in Fig.~\ref{fig:td-dipole-moment} (a), we find first regular Rabi-oscillations up to the maximum at $t=5$~ps and around $t=10$~ps, we find the neck-like feature ~\cite{fuks2014} typical for Rabi-oscillations. In Fig.~\ref{fig:td-dipole-moment} (b), we show the purity $\gamma$ in dashed black lines. The purity $\gamma$, which is a measure for the separability of the many-body wave function into a product of an electronic and a photon wave function. We find that $\gamma$ is close to $1$ up to $t=5$~ps, which means that the many-body wave function is close to a factorizable state. After $t=5$~ps, $\gamma$ deviates strongly from $1$ and the system is not factorizable anymore. This dynamical build-up of correlation has also an effect on the non-classicality of the light-field visible in the Mandel $Q$-parameter shown in Fig.~\ref{fig:td-dipole-moment} (b) in solid black lines. While initially $Q\approx0$ that indicates the coherent statistics of the photon mode, after $t=5$~ps also this observable deviates from $0$ and nonclassicality shows up. From Fig.~\ref{fig:td-dipole-moment} (c), where we plot the photon number, we see that until $t=5$~ps a photon is absorbed that is later re-emitted and after $t=15$~ps, we again observe photon absorption processes.
\begin{figure}[h] 
\centerline{\includegraphics[width=0.5\textwidth]{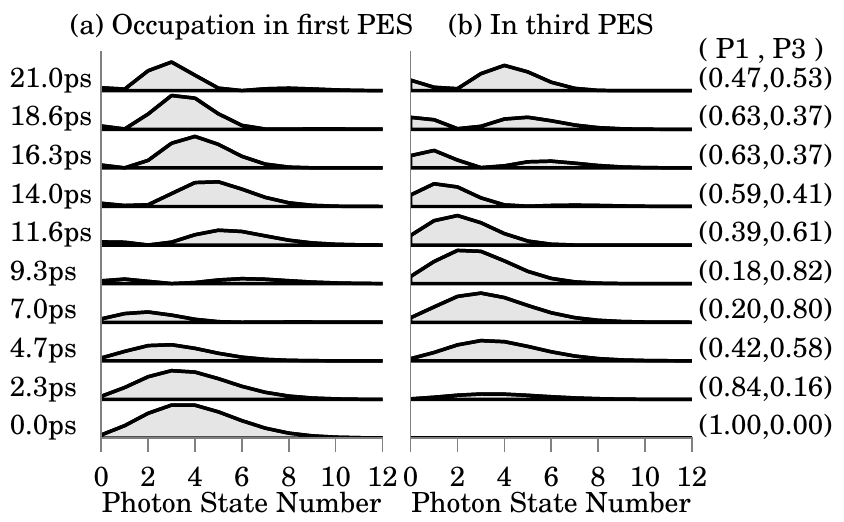}}
\caption{Photon population in the first and third PES for the case discussed in Fig.~\ref{fig:td-dipole-moment}. }
\label{fig:td-population}
\end{figure}
In the following, we {analyze} this dynamics of the correlated electron-photon problem in terms of population in the cavity Born-Oppenheimer surfaces calculated in Fig.~\ref{fig:qedft-fp-bo-surfaces} (a). In Fig.~\ref{fig:td-population}, we show the occupation of the photon number states in the first {cavity} {PES} in (a) and the third {cavity} {PES} in (b). The values (P$1$,P$3$) give the population of the first {cavity} {PES} and the third {cavity} {PES}, respectively. All other {cavity} {PES} have populations which are an order of magnitude smaller, since P$1+$P$3$ is close to $1$ for all times. In Fig.~\ref{fig:td-population} (a), we find that at the initial time $t=0$~ps, the first {cavity} {PES} is populated with a photon state, which has a coherent distribution with $\left<\hat{a}^\dagger_\alpha \hat{a}_\alpha\right>=4$, which is in agreement with our initial condition. During the time propagation, we observe a transfer of population from the first {cavity} {PES} to the third {cavity} {PES}. In the first {cavity} {PES}, we see until $t=9.3$~ps a depletion of population, while in the third {cavity} {PES} (Fig.~\ref{fig:td-population} (b)), we observe an increase of the population. After this time, the population is again transferred back from the third {cavity} {PES} to the first {cavity} {PES} (Rabi oscillation). However, not only the amplitude of the population is changing, but also the center of the wave packets. In principle, if the same photon state would be populated in the two different {cavity} {PES}, the system could still be factorizable. For small times, up to $t=5$~ps the center of the wave packet in the first {cavity} {PES} remains close to its initial value. Later it changes to smaller photon numbers, which indicates photon absorption. We can conclude that the dynamics of the many-body system is dominated by the population transfer from the first {cavity} {PES} to the third {cavity} {PES} and vice versa. While for this example, a good approximate description may be a two-surface approximation reminiscent of the Rabi model~\cite{braak2011}, we expect a different behavior for more complex cavity Born-Oppenheimer surfaces~e.g.~in many-electron problems, multi-photon modes, or strong-coupling situations.

\subsection{Light-Matter coupling via vibrational excitation}

The second system that we analyze is the  Shin-Metiu model~\cite{shin1995,shin1996} {coupled to cavity photons}. {Without coupling to photon modes,} this system exhibits a conical intersection between Born-Oppenheimer surfaces and has been analyzed heavily in the context of correlated electron-nuclear dynamics~\cite{albareda2014}, exact forces in non-adiabatic charge transfer~\cite{agostini2015}, or nonadiabatic effects in quantum reactive scattering~\cite{peng2014}, to mention a few. In our case, we place the system, consisting of three nuclei and a single electron into a optical cavity, where it is coupled to a single mode that is in resonance with the first vibrational excitation. The outer two nuclei are fixed and the free electron and the nuclei are restricted to one-dimension. The model is schematically depicted in Fig.~\ref{fig:shin-model}.
\begin{figure}[ht]
\centerline{\includegraphics[width=0.3\textwidth]{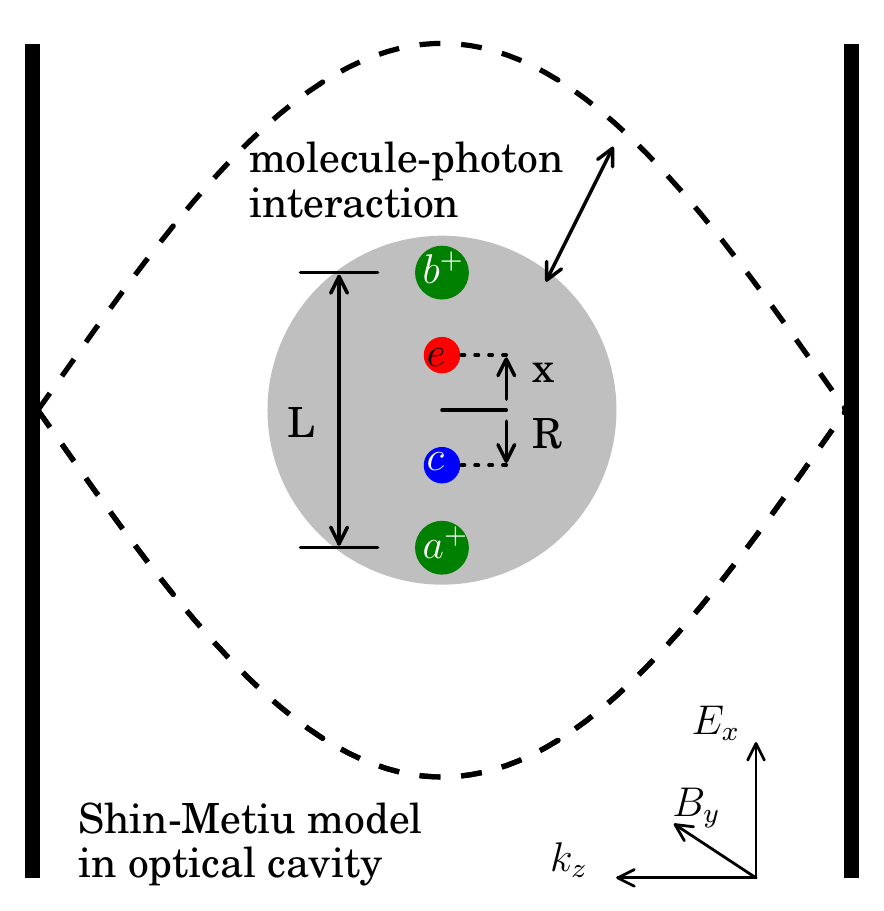}}
\caption{Molecule in an optical cavity. The molecule is modeled by the Shin-Metiu model~\cite{shin1995,shin1996}
that consists of three nuclei and a single electron. Two of the nuclei are frozen at position $L/2$ and $-L/2$, respectively.}
\label{fig:shin-model} 
\end{figure}
The Hamiltonian of such a system is given by~\cite{shin1995,shin1996}
\begin{align}
\hat{H} = -\frac{\hbar}{2M} \frac{\partial^2}{\partial R^2} + \hat{H}_e + \hat{H}_p + \hat{H}_{pe} + \hat{H}_{pn} + \hat{H}_{pen},
\end{align}
where $\hat{H}_p$, $\hat{H}_{pe}$, $\hat{H}_{pn}$, and $\hat{H}_{pen}$ are given by Eqns.~\ref{eq:correlated-photon-hamiltonian},~\ref{eq:correlated-electron-photon-hamiltonian},~\ref{eq:correlated-photon-nuclear-hamiltonian},~\ref{eq:correlated-electron-nuclear-photon-hamiltonian}, respectively. The electronic Hamiltonian {reads}
\begin{align}
\hat{H}_e = -\frac{\hbar}{2M} \frac{\partial^2}{\partial r^2} + V_n(R) + V_e(r,R),
\end{align}
where $V_n(R)$ is the Coulomb interaction of the free nuclei with the two fixed nuclei, $r$ is the electronic coordinate and $R$ the nuclear coordinate. $V_e(r,R)$ is given by
\begin{align}
V_e(r,R) = Ze^2 \text{erf}\left(\left(r-R\right)/R_c\right)/\left(r-R\right),
\end{align}
where $\text{erf}$ describes the error-function. We fix the nuclear mass $M$ to the mass of a hydrogen atom and the length $L=10\AA$. Furthermore, we use the  dipole operators $X_e=-er$ and $X_n=eR$. Further $R_c$ can be used to tune the energy difference $\Delta$ between the ground-state and the first-excited state  potential energy surface. For the cavity Shin-Metiu {model}, we {represent} the electron on a grid of dimension $N_r=140$ with $\Delta r=0.4233\AA$, and the nuclear coordinate on a grid of dimension $N_R=280$ with $\Delta R=0.0265\AA$, while the photon {wave function is expanded} in the photon number eigenbasis, where the mode can host up to $81$ photons in the photon mode. To get first insights on how the light-matter coupling is capable of changing the chemical landscape of the system, in Fig.~\ref{fig:shin-results1}, we calculate the ordinary PES surfaces of Eq.~\ref{eq:CBOA-surfaces} for the case of $q_\alpha=0$. The solid red line shows the ground-state energy surface, while the blue line shows the excited state energy surface for $R_c=1.5\AA$ with $\hbar\omega_\alpha=72.5$~meV and $R_c=1.75\AA$ with $\hbar\omega_\alpha=69.3$~meV. In both examples, the photon frequencies $\omega_\alpha$ correspond to the first vibrational transition of the exact bare Hamiltonian. 
\begin{figure}[ht]
\centerline{\includegraphics[width=0.5\textwidth]{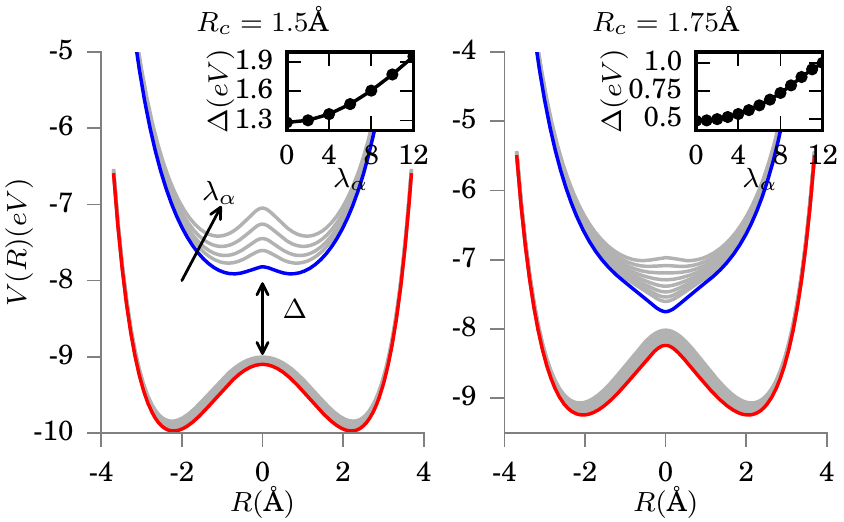}}
\caption{Potential energy surfaces in the cavity Born-Oppenheimer approximation for the Shin-Metiu model. Increasing matter-photon coupling strength opens the gap $\Delta$ between the ground-state {cavity} PES and the first-excited {cavity} PES. Both plots are using parameters as in Ref.~\cite{shin1995} and are evaluated at $q_{\alpha}=0$.}
\label{fig:shin-results1}
\end{figure}
{Next}, we tune the matter-photon coupling strength $\lambda_\alpha$ from the weak-coupling regime to the strong-coupling regime. {The corresponding cavity PES are shown in grey in Fig.~\ref{fig:shin-results1}}. The inset in the figures shows the energy gap $\Delta$ depending on the matter-photon coupling strength $\lambda_\alpha$. In the left figure, we choose the value $R_c=1.5\AA$ and in the case of $\lambda_\alpha = 0$, we find well separated cavity Born-Oppenheimer surfaces. The matter-photon coupling (chosen here from $\lambda_\alpha=0$ to $\lambda_\alpha=82.55$~eV$^{1/2}$/nm with a Rabi-splitting $\Omega_R=\left(E_5-E_3\right)/\hbar\omega_\alpha)=43.81\%$) opens the gap significantly, as shown in the inset. Additionally, for $R_c=1.5\AA$, we find that the double well structure visible in the first-excited state becomes more pronounced for stronger light-matter coupling. The right figure shows the results for $R_c=1.75\AA$, where in the field-free case a much narrower gap $\Delta$ is found. Introducing the matter-photon coupling in the system from $\lambda_\alpha= 0$ to $\lambda_\alpha= 84.48$~eV$^{1/2}$/nm with $\Omega_R=64.04\%$, also opens the gap significantly and we find a similar qualitative behavior as in the previous example with the {notable} difference, that we observe {in the present example} a similar single-well to double well transition {but now} in the first-excited state. However, {since} we restricted ourselves to a specific cut in the full two-dimensional cavity Born-Oppenheimer surface by choosing $q_\alpha=0$, Fig.~\ref{fig:shin-results1} does not show the full picture.   
\begin{figure}[ht] 
\centerline{\includegraphics[width=0.5\textwidth]{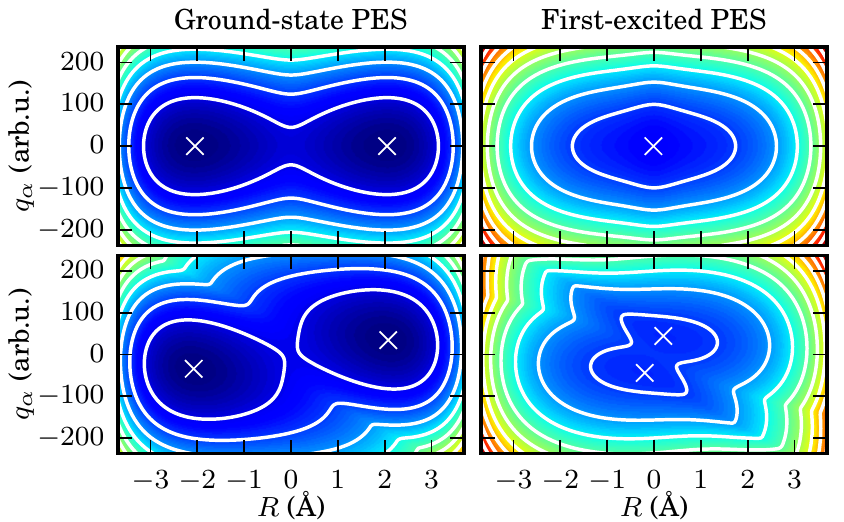}}
\caption{Two-dimensional ground-state and first-excited state potential energy surfaces in the cavity Born-Oppenheimer approximation for the Shin-Metiu model in the case of $\lambda_\alpha=0$ (upper panel) and strong-coupling $\lambda_\alpha=79.20$~eV$^{1/2}$/nm. (lower panel) with  $R_c=1.75\AA$. High-energy regions are plotted by red color, while low-energy regions are plotted by blue color. The crosses denote the minima of the surfaces.}
\label{fig:shin-results2}
\end{figure}
Therefore, in Fig.~\ref{fig:shin-results2}, we show the full two-dimensional cavity  PES for $R_c=1.75\AA$. In the figure, the x-axis show the nuclear degree of freedom ($R$), while the y-axis shows the photonic degree of freedom $q_\alpha$. In the case of $\lambda_\alpha=0$, that is the upper panel in the figure, we find that the photonic degree of freedom introduces anharmonicity into the surface.  We also indicate the minima in the surfaces by white crosses. In agreement with Fig.~\ref{fig:shin-results1}, we find a double {minimum} for the ground-state {cavity} PES and a single {minimum} for the excited state {cavity} PES. In the case of strong-coupling that is shown in the lower panel of the figure, we observe new emerging normal modes. These new normal modes are caused by the entanglement of the matter and photon degrees of freedom and are manifest in the displacement of the minima out of the equilibrium positions. In the first-excited state surface in strong coupling, we also observe a single-well to double-well transition, as observed in the coupling to the electronic excitation and discussed in the first part of this work. Here, we find that now two minima appear in the first-excited state surface. If we adopt an adiabatic picture we can conclude that now two new reaction pathways are possible from the first excited state surface to the ground-state surface.\\
To conclude, we have seen how the photonic degrees of freedom alter considerably chemical properties in a model system containing electronic, nuclear and photonic degrees of freedom. We have identified the change of {traditional} Born-Oppenheimer surfaces, gap opening, and transitions from single well structures to double-well structures in the first-excited state surface from first principles. The gap opening can be connected to recent experiments~\cite{george2015}, where a reduction in chemical activity has been observed for vibrational strong coupling.
 
\section{Summary and Outlook}
In this paper, we introduced the concept of the cavity Born-Oppenheimer approximation for electron-nuclear-photon systems. We used the cavity Born-Oppenheimer {approximation}  to analyze the ground-state transition in the system that emerges in the strong-coupling limit. During this transition the ground-state electron density is split and the ground-state {cavity} PES obtains a double well structure featuring finite displacements of the photon coordinate. Furthermore, we illustrated for a time-dependent situation with a factorizable initial state, how the complex correlated electron-photon dynamics can be interpreted by an underlying back-and-forth photon population transfer from the ground-state {cavity} PES to an excited-state {cavity} PES. In the last section, we have demonstrated how this transition can also appear in case of  strong-coupling {and vibrational resonance}. Here, we find that the first-excited state surface can obtain a double well structure leading to new reaction pathways in an adiabatic picture. {In} future studies towards a full ab-initio description for cavity light-matter systems, where solving the electronic Schr\"odinger equation of Eq.~\ref{eq:electron-schrodinger} by exact diagonalization is not feasible, the density-functional theory for electron-photon systems (QEDFT) can be used~\cite{tokatly2013,ruggenthaler2014}. The discussed methods can be still improved, e.g.~{along} the lines of a more accurate factorization method {such as} the exact factorization~\cite{abedi2010,abedi2012,eich2016} known for electron-nuclear problems, or trajectory based methods~\cite{albareda2014,albareda2015} can be applied to simulate such systems dynamically. This work has direct implications on more complex correlated matter-photon {problems} that can be approximately solved employing the cavity Born-Oppenheimer {approximation} to better understand complex correlated light-matter coupled systems.

\section{Acknowledgements}
We would like to thank C. Sch\"afer for a careful reading of the manuscript, and  {MR acknowledges insightful discussions with F.G.\ Eich}. We acknowledge financial support from the European Research Council (ERC-2015-AdG-694097), Grupos Consolidados (IT578-13), by the European Union's H2020 program under GA no.676580 (NOMAD), COST Action MP1306 (EUSpec) and the Austrian Science {Fund} (FWF P25739-N27).

\bibliographystyle{apsrev4-1}
\bibliography{01_light_matter_coupling} 

\end{document}